\documentclass[12pt]{article}
\pdfoutput=1

\newlength{\abstractwidth}
\usepackage[height=8.85in,width=6.45in]{geometry}
\renewcommand{\baselinestretch}{1.05}

\usepackage{amsmath}
\usepackage{amsfonts}
\usepackage{amssymb}
\usepackage{latexsym}
\usepackage{color}
\usepackage{dsfont}
\usepackage{graphicx}
\usepackage{microtype}
\usepackage[svgnames,dvipsnames]{xcolor}
\usepackage[colorlinks,citecolor=DarkGreen,linkcolor=FireBrick,urlcolor=FireBrick,linktocpage,unicode,psdextra]{hyperref}

\definecolor{darkred}{rgb}{0.8,0.1,0.1}

\renewcommand{\thefootnote}{\fnsymbol{footnote}}
\renewcommand{\thanks}[1]{\footnote{#1}}
\newcommand{\starttext}{
\setcounter{footnote}{0}
\renewcommand{\thefootnote}{\arabic{footnote}}}
\numberwithin{equation}{section}
\newcommand{\bea}{\begin{eqnarray}}
\newcommand{\eea}{\end{eqnarray}}
\newcommand{\be}{\begin{eqnarray}}
\newcommand{\ee}{\end{eqnarray}}

\def\cG{{\cal G}}

\def\cO{{\cal O}}

\def\cY{{\cal Y}}

\def\ZZ{{\mathbb Z}}

\def\Im{{\rm Im\,}}
\def\tr{{\rm tr}}

\def\det{{\rm det\,}}

\def\no{\nonumber}

\makeatletter\def\l@subsubsection#1#2{}%
\makeatother

\begin{document}

\begin{titlepage}

\begin{flushright}
KYUSHU-HET-373
\end{flushright}

\vskip 1.4cm

\begin{center}

{\Large \bfseries Fermion Hierarchies from Fricke Modular Symmetry}

\vskip 1.4cm

Takafumi Kai$^1$,
Justin Kaidi$^{1,2,3}$,
Katsumi Kina$^{4}$,
Hajime Otsuka$^{1,3}$
\vskip 1cm

\begin{tabular}{c p{0.78\textwidth}}
1 & Department of Physics, Kyushu University, Fukuoka 819-0395, Japan \\
2 & Institute for Advanced Study, Kyushu University, Fukuoka 819-0395, Japan\\
3 & Quantum and Spacetime Research Institute (QuaSR), Kyushu University, Fukuoka 819-0395, Japan\\
4 & Graduate School of Mathematics, Kyushu University, Fukuoka 819-0395, Japan
\end{tabular}

\vskip 1.4cm

\end{center}

\begin{abstract}
We show how Fricke modular symmetry can generate realistic fermion hierarchies and quark CP violation from a single modulus in a supersymmetric effective theory. Matter fields are labelled by the two one-dimensional characters of $\Gamma_0^+(N)$ that are trivial on $\Gamma_0(N)$, and interactions use complete spaces of cusp-regular holomorphic modular forms. At level three, the Fricke extension enlarges an interior stabilizer from order three to order six, allowing its parity to enforce three additional powers of the displacement $t$. An explicit assignment gives quark mass powers $(7,3,0)$ and $(5,4,2)$ and CKM powers $(1,2,3)$. The leading textures conserve CP, but an exact relation among the modular generators produces an appreciable CKM phase at finite displacement. A fit at $|t|\simeq0.306$ reproduces the quark and lepton reference values with order one coefficients in a fixed normalization. We also relate the Fricke double-coset algebra to coset non-invertible symmetry; on the coefficient spaces used here its action reduces to ordinary Fricke parity.
\end{abstract}
\end{titlepage}
\pagenumbering{roman}
\tableofcontents
\clearpage
\pagenumbering{arabic}
\starttext

\section{Introduction}
\label{sec:introduction}

The pattern of fermion masses and mixing remains one of the central puzzles of particle physics. Quarks and charged leptons have widely separated masses, and quark mixing is small and hierarchical. The Standard Model accommodates these features through independent Yukawa couplings, but does not explain their origin. Neutrino oscillations add a different pattern, with large lepton mixing angles. A flavor symmetry can relate these couplings and force some of them to vanish in a symmetric limit. The challenge is then to explain how symmetry breaking generates the observed hierarchies, and to identify which features follow from symmetry rather than from a choice of numerical coefficients.

Modular flavor provides a natural setting for this question. In such models, the Yukawa couplings are taken to be holomorphic functions of a complex scalar modulus $\tau$, with transformation laws fixed by a modular symmetry. Near a value of $\tau$ that preserves part of this symmetry, the couplings can begin at different powers of the displacement from that point. A single small displacement can therefore generate several hierarchies, with their powers determined by the transformation properties of the fields and coefficient functions \cite{Feruglio:2017spp,Feruglio:2021dte,Novichkov:2021evw}. This mechanism makes it possible to obtain small fermion masses without assigning equally small coefficients to their interactions.\footnote{For a review of modular flavor models, see \cite{Ding:2023htn}. Related studies include universal scaling near fixed points \cite{Feruglio:2023mii}, explicit quark and lepton hierarchies \cite{Kikuchi:2023cap,Abe:2023qmr}, and the consistent combination of modular and CP symmetries, with CP broken only by the modulus \cite{Novichkov:2019sqv}. 
Yukawa couplings described by modular forms of congruence subgroups, including $\Gamma_0(3)$, also arise in string compactifications~\cite{Ishiguro:2024xph}.}

In this paper we study the additional constraints supplied by Fricke modular symmetry. The Fricke group $\Gamma_0^+(N)$ extends $\Gamma_0(N)$ by $W_N:\tau\mapsto-1/(N\tau)$. We use its two one-dimensional characters, denoted $+$ and $-$, which are trivial on $\Gamma_0(N)$ and distinguished by their Fricke parity. Interactions are ordinary invariant products of local chiral multiplets and scalar modular forms. Their coefficient spaces are the corresponding Fricke eigenspaces of $M_k(\Gamma_0(N))$.

The usefulness of the Fricke extension becomes apparent at its interior elliptic points. The requirement that the added Fricke component supply a stabilizer of order greater than two singles out levels two and three. At level three, the relevant point and a local displacement are
\bea
\tau_*={1\over2}+{i\over2\sqrt3}~,
 \qquad t={\tau-\tau_*\over\tau-\bar\tau_*}~.
\eea
The transformation $TW_3$ rotates $t$ by $e^{-i\pi/3}$ and distinguishes Taylor powers modulo six. Its square, an ordinary transformation in $\Gamma_0(3)$, distinguishes them only modulo three. Fricke parity can therefore exclude a lower-order term and enforce three additional powers of the displacement. For example, the two weight-eight character spaces begin at $t^2$ and $t^5$. We determine the complete coefficient spaces and prove that every nonempty space attains its least symmetry-allowed order.

An explicit matter assignment then gives the generic quark textures
\bea\label{eq:introtextures}
Y_u\sim\begin{pmatrix}t^4&t^3&t\\0&t^2&1\\0&0&t^4\end{pmatrix}~,
 \qquad Y_d\sim\begin{pmatrix}t^5&t^4&t^2\\0&t^4&t^2\\0&t^4&t^2\end{pmatrix}~.
\eea
The zeros follow from empty holomorphic coefficient spaces. The leading minors and determinants yield\footnote{Hierarchies near modular fixed points and realistic modular flavor fits have been studied extensively, see e.g. \cite{Novichkov:2021evw,Liu:2021gwa,Petcov:2022fjf,Chen:2025tby}; here the Fricke extension provides an interior order-six fixed point and a particularly simple realization using only two one-dimensional characters.}
\bea\label{eq:introhierarchies}
y_u:y_c:y_t&\sim&|t|^7:|t|^3:1~,
 \no\\
 y_d:y_s:y_b&\sim&|t|^5:|t|^4:|t|^2~,
 \no\\
 |V_{us}|:|V_{cb}|:|V_{ub}|&\sim&|t|:|t|^2:|t|^3~.
\eea
These powers hold for generic coefficients and persist under regular symmetry-compatible matter metrics. Fricke parity is essential: removing it while keeping the weights fixed would change the generic up-sector mass powers to $(|t|,1,1)$.

CP violation probes information beyond these leading powers. With real coefficients, all phases in the leading quark textures can be removed by field redefinitions. The exact modular forms, however, obey a ring relation that generates a nontrivial relative phase at finite displacement. At $|t_0|\simeq0.306$, we find a fit reproducing six quark Yukawas and all four CKM parameters with every allowed basis element present. All quark coefficients, including both overall normalizations, are real and have magnitudes of order one in a basis fixed at the symmetry point. The vev of $\tau$ is the only CP-breaking input. The same vacuum also reproduces the charged-lepton and neutrino reference values through a type-I seesaw with coefficients of order one.

We work in a rigid $\mathcal N=1$ effective theory with cusp-regular flavor coefficients. Modulus stabilization and supersymmetry breaking are left for future work. We will also explain a relation between the Fricke double-coset algebra and coset non-invertible symmetry. On the coefficient spaces used here this action reduces to ordinary Fricke parity, so the local flavor construction requires only invertible symmetry.

\paragraph{Organization:} Section \ref{sec:fricke} introduces the Fricke symmetry at general level, then identifies the elliptic points and coefficient spaces relevant to hierarchies. Section \ref{sec:quarks} derives the quark hierarchies, explains spontaneous CP violation, and presents a numerical fit. Section \ref{sec:leptons} treats the charged leptons and seesaw mechanism. Appendix \ref{app:modularspaces} proves completeness of the modular-form bases, Appendix \ref{app:normalization} explains their normalization and local effective-theory interpretation, and Appendix \ref{app:numericaldata} collects the full numerical data. Appendix \ref{app:anomalies} checks mixed modular-gauge anomalies, and Appendix \ref{app:doublecosets} derives the Fricke double-coset algebra.

\section{Fricke modular flavor symmetry}
\label{sec:fricke}

We consider a four-dimensional $\mathcal N=1$ effective theory with a dynamical chiral multiplet whose scalar component $\tau$ takes values in the upper half-plane. We use the same symbol $\tau$ for the corresponding chiral superfield. The flavor symmetry is taken to be the Fricke group
\bea\label{eq:Frickegroup}
\Gamma^+_0(N):=\langle\Gamma_0(N),W_N\rangle~,
 \hspace{0.4 in}
 W_N:\,\tau\longmapsto-{1\over N\tau}~,
 \hspace{0.4 in} N\in\ZZ_{>1}~,
\eea
where $\Gamma_0(N)$ consists of integral determinant-one matrices whose lower-left entry is divisible by $N$. We work projectively, so that $-\mathds{1}$ acts trivially. The Fricke involution normalizes $\Gamma_0(N)$, and the extension has two components, $\Gamma_0(N)$ and $\Gamma_0(N)W_N$.

The matter fields are taken to carry one-dimensional characters of this ordinary modular group. We restrict to the two characters that are trivial on $\Gamma_0(N)$,
\bea\label{eq:Frickecharacters}
\chi_\epsilon(\gamma)=\left\{ \begin{matrix} \,\,1 & \gamma \in \Gamma_0(N)
 \\
\,\, \epsilon & \hspace{0.3 in}\gamma \in \Gamma_0(N) W_N
 \end{matrix} \right.
 ~,
\eea
where $\epsilon=\pm1$. Their tensor product is $\chi_{\epsilon_1}\chi_{\epsilon_2}=\chi_{\epsilon_1\epsilon_2}$. We label the corresponding representations simply by $+$ and $-$, so a matter multiplet is specified by a modular weight $k_\Phi$ and a sign $\epsilon_\Phi$. These characters give the simplest scalar coefficient spaces; the relation to the larger double-coset algebra will be discussed at the end of this section.

\subsection{Supersymmetric effective theory}

For a positive-determinant matrix $\gamma=\left(\begin{smallmatrix}a&b\\c&d\end{smallmatrix}\right)$, define
\bea\label{eq:automorphyfactor}
\gamma\tau={a\tau+b\over c\tau+d}~,
 \hspace{0.4 in} j_\gamma(\tau)={c\tau+d\over \sqrt{\det\gamma}}~.
\eea
Thus $j_{W_N}(\tau)=\sqrt N\tau$ when $W_N$ is represented by $\left(\begin{smallmatrix}0&-1\\N&0\end{smallmatrix}\right)$. The transformation law of a chiral multiplet is
\bea\label{eq:mattertransformation}
\Phi\longmapsto\chi_{\epsilon_\Phi}(\gamma)j_\gamma(\tau)^{-k_\Phi}\Phi~.
\eea
All weights are even, consistently with the projective convention and the absence of an additional multiplier system. Matter weights need not be positive.

Following the standard supersymmetric modular flavor construction \cite{Feruglio:2017spp,Feruglio:2021dte}, we take
\bea\label{eq:kahlerpotential}\label{eq:yukawasuperpotential}
K&=&-\Lambda_K^2\log[-i(\tau-\tau^\dagger)]+\sum_\Phi{\Phi^\dagger e^{2V}\Phi\over [-i(\tau-\tau^\dagger)]^{k_\Phi}}~,\no\\
 W_Y&=&\sum_{i,j}\left[\mathcal Y^u_{ij}(\tau)U_i^cQ_jH_u+\mathcal Y^d_{ij}(\tau)D_i^cQ_jH_d+\mathcal Y^e_{ij}(\tau)E_i^cL_jH_d\right]~.
\eea
Here $\Lambda_K$ sets the normalization of the modulus kinetic term, and the sum in $K$ includes any gauge-singlet neutrino multiplets introduced below. Gauge indices in the superpotential are contracted in the usual way. We assume matter parity, with matter multiplets odd and Higgs and modulus multiplets even. We impose generalized CP on the full effective theory, taking the gauge, Higgs, and supersymmetry-breaking sectors to be CP invariant and all superpotential coefficients real in the CP basis \cite{Chen:2014tpa,Novichkov:2019sqv,Kobayashi:2025wty}. CP is then broken only by the vev of $\tau$, as described in Section \ref{sec:spontaneousCP}.

Since $2\Im(\gamma\tau)=2\Im\tau/|j_\gamma|^2$, the matter kinetic terms are invariant. The modulus term changes by the K\"ahler transformation $\Lambda_K^2\log j_\gamma+\mathrm{c.c.}$, which does not affect the rigid supersymmetric action. We require the superpotential to be invariant and of modular weight zero.\footnote{In supergravity, the K\"ahler transformation must be accompanied by $W\mapsto j_\gamma^{-\Lambda_K^2/M_{\rm Pl}^2}W$. The assignments below use rigid supersymmetry and are not supergravity weight assignments.} Consequently, the coefficient of $F_i^cQ_jH_q$ has weight and character
\bea
\label{eq:coefficientassignments}
K^q_{ij}=k_{F_i^c}+k_{Q_j}+k_{H_q}~,
 \hspace{0.4 in} \epsilon^q_{ij}=\epsilon_{F_i^c}\epsilon_{Q_j}\epsilon_{H_q}~.
\eea
The coefficient transforms with $j_\gamma^{K^q_{ij}}\chi_{\epsilon^q_{ij}}(\gamma)$, so each displayed product is an explicit invariant local interaction.

Holomorphy in the chiral modulus is enforced by supersymmetry. We further require the Yukawa coefficients to be regular at every cusp. This is an additional assumption on the effective flavor couplings: holomorphy in the upper half-plane alone would also permit functions with cusp poles. With this assumption, a negative coefficient weight gives an exact superpotential zero, while at weight zero only an even constant is available.

At a fixed vacuum $\tau_0$, canonical normalization with the metric in (\ref{eq:kahlerpotential}) gives
\bea
\label{eq:canonicalyukawa}
\Phi^{\rm can}=(2\Im\tau_0)^{-k_\Phi/2}\Phi~,
 \hspace{0.4 in} Y^q_{ij}=(2\Im\tau_0)^{K^q_{ij}/2}\mathcal Y^q_{ij}(\tau_0)~.
\eea
These factors are finite and nonzero at an interior fixed point and preserve the leading order of every matrix entry, but they must  be included in a numerical evaluation. 
Modular symmetry does not uniquely determine the K\"ahler potential,
and additional invariant terms can affect the physical masses
and mixing parameters \cite{Chen:2019ewa}. 
Our numerical analysis uses the minimal diagonal metric in (\ref{eq:kahlerpotential}).

\subsection{Complete coefficient spaces}
\label{sec:modularcoefficients}

We use the determinant-normalized slash operation
\bea
(f|_k\gamma)(\tau)=j_\gamma(\tau)^{-k}f(\gamma\tau)~.
\eea
Let $M_k(\Gamma_0(N))$ denote the weight-$k$ forms holomorphic in the upper half-plane and at every cusp, with trivial character on $\Gamma_0(N)$. Fricke conjugation preserves $\Gamma_0(N)$, and $W_N$ is an involution on the even-weight spaces. The complete coefficient spaces are
\bea\label{eq:Frickecoefficientspaces}
M_k^\epsilon(N)=\{f\in M_k(\Gamma_0(N)):f|_kW_N=\epsilon f\}~,
 \qquad M_k(\Gamma_0(N))=M_k^+(N)\oplus M_k^-(N)~.
\eea
Explicit elements of these spaces can be built from the Eisenstein series
\bea\label{eq:Eisensteinseries}
E_k(\tau)=1-{2k\over B_k}\sum_{n\geq1}\sigma_{k-1}(n)q^n~,
 \qquad q=e^{2\pi i\tau}~,
\eea
where $B_k$ is a Bernoulli number and $\sigma_{k-1}(n)$ is the sum of the $(k-1)$-st powers of the positive divisors of $n$. For even $k\geq4$, the Fricke slash operation exchanges $E_k(\tau)$ and $N^{k/2}E_k(N\tau)$, so
\bea
\label{eq:FrickeEisensteinexamples}
E_k(\tau)+\epsilon\, N^{k/2}E_k(N\tau)\in M_k^\epsilon(N)~.
\eea
At weight two, $E_2$ itself is quasimodular, but the combination $E_2(\tau)-NE_2(N\tau)$ is holomorphic and Fricke odd, and hence belongs to $M_2^-(N)$. These examples make the two Fricke parities explicit. The complete spaces may also contain additional independent Eisenstein series and cusp forms; Appendix \ref{app:modularspaces} constructs complete bases at the levels used below.

Each basis element has its own superpotential coefficient, which is real in the CP basis specified below. We will include the full space rather than select a particular Eisenstein series or product of generators. The resulting suppressions therefore survive all coefficient functions permitted by the symmetry.

\subsection{Interior fixed points and local expansions}
\label{sec:fixedpoints}

To generate hierarchical couplings, we place the modulus near an interior fixed point and use its residual symmetry to constrain the local Taylor expansion. The Fricke involution fixes $\tau=i/\sqrt N$ at every level. However, in the local coordinate $t_F=(\tau-i/\sqrt N)/(\tau+i/\sqrt N)$ it acts by $t_F\mapsto-t_F$, and hence this residual symmetry distinguishes Taylor powers only modulo two; by itself it cannot enforce a zero beyond first order. We therefore seek elliptic transformations of higher order in the Fricke component.

A useful representative is
\bea\label{eq:gNdefinition}
g_N=TW_N:\,\,\,\,\,\tau\longmapsto1-{1\over N\tau}~, \hspace{0.4 in} T:\,\,\tau\longmapsto\tau+1~.
\eea
Its determinant-normalized trace is $\sqrt N$, so $g_N$ is elliptic for $N=2,3$, parabolic for $N=4$, and hyperbolic for $N>4$. More generally, for $\gamma=\left(\begin{smallmatrix}a&b\\Nc&d\end{smallmatrix}\right)\in\Gamma_0(N)$, the trace of the determinant-one representative of $\gamma W_N$ is
\bea\label{eq:Frickecomponenttrace}
{\tr(\gamma W_N)\over \sqrt{\det(\gamma W_N)}}=\sqrt N(b-c)~.
\eea
For $N\geq4$, ellipticity requires $b=c$, giving trace zero and projective order two. Thus $N=2,3$ are the only levels $N>1$ at which the Fricke component supplies an elliptic transformation of order greater than two.\footnote{This statement concerns the additional Fricke transformations; it does not exclude order-three elliptic points in the integral subgroup $\Gamma_0(N)$ at higher level.} We therefore focus on these levels.

For $N=2,3$, the group is generated by $T$ and $W_N$ and has signature $(2,4,\infty)$ or $(2,6,\infty)$, respectively \cite{Doran:2013npa}. Concretely, there are two inequivalent elliptic points in the upper half-plane. At level two their residual symmetries are $\ZZ_2$ and $\ZZ_4$, while at level three they are $\ZZ_2$ and $\ZZ_6$. A vacuum at either point preserves the corresponding finite symmetry, which constrains the Taylor powers of the couplings around it. The final entry $\infty$ denotes the cusp: a boundary degeneration with a parabolic, infinite-order stabilizer, rather than another interior fixed point. 

The $\ZZ_2$ point is represented by the aforementioned $i/\sqrt N$, while the higher-order points are
\bea\label{eq:ellipticpoint}
\tau_{N*}={1\over 2}+{i\over 2}\sqrt{{4\over N}-1}~,
\eea
which are fixed by $g_N$. In particular, we have 
\bea
 \hspace{0.4 in} \tau_{2*}={1+i\over 2}~,
 \hspace{0.4 in} \tau_{3*}={1\over 2}+{i\over 2\sqrt3}~.
\eea
Let us now introduce the holomorphic disc coordinate
\bea\label{eq:discparameter}
t_N={\tau-\tau_{N*}\over \tau-\bar\tau_{N*}}~,
 \hspace{0.4 in} \tau(t_N)={\tau_{N*}-t_N\bar\tau_{N*}\over 1-t_N}~.
\eea
Direct substitution gives
\bea\label{eq:discrotation}
t_N(g_N\tau)=e^{-i\pi/N}t_N(\tau)~.
\eea
It is this fact which will allow us to constrain the Taylor expansions of coefficient functions and generate fermion hierarchies.

Concretely, for a function $f\in M_k^\epsilon(N)$, remove the varying automorphy factor by defining
\bea\label{eq:covariantdiscframe}
\widehat f(t_N)=(1-t_N)^{-k}f(\tau(t_N))~.
\eea
The prefactor is nonzero at $t_N=0$ and does not change the order of a zero. Since $\sqrt N\tau_{N*}=e^{i\pi/(2N)}$, the transformation under $g_N$ becomes
\bea
\widehat f(e^{-i\pi/N}t_N)=\epsilon\, e^{i\pi k/(2N)}\widehat f(t_N)~.
\eea
Writing $\epsilon=(-1)^p$, a nonzero Taylor coefficient at order $n$ consequently requires
\bea\label{eq:fixedpointcongruence}
n\equiv Np-{k\over 2}\pmod{2N}~.
\eea
At these levels, the complete bases attain the least nonnegative solution for every nonempty space in (\ref{eq:Frickecoefficientspaces}), as shown in Appendix \ref{app:modularspaces}. Thus a generic form has
\bea\label{eq:genericexpansion}
\widehat f(t_N)=t_N^{n_\epsilon(k)}\left(c_0+c_1t_N^{2N}+c_2t_N^{4N}+\cdots\right)~,
 \hspace{0.25 in} n_\epsilon(k)=\left[Np-{k\over 2}\right]_{2N}~,
\eea
where  $c_0\neq0$ and $[\cdot]_m$ is the least nonnegative residue modulo $m$. The resulting orders are
\bea\label{eq:ordertables}
\begin{array}{c|cc}
 k\bmod8&n_+(k)&n_-(k)\\\hline
 0&0&2\\2&3&1\\4&2&0\\6&1&3
\end{array}
\hspace{0.2 in}(N=2)~,
\hspace{0.2 in}
\begin{array}{c|cc}
 k\bmod12&n_+(k)&n_-(k)\\\hline
 0&0&3\\2&5&2\\4&4&1\\6&3&0\\8&2&5\\10&1&4
\end{array}
\hspace{0.2 in}(N=3)~.\hspace{0.1 in}
\eea

The importance of the Fricke extension is particularly transparent at level three. The square $g_3^2$ is an integral order-three transformation in $\Gamma_0(3)$, whereas $g_3$ has order six. The first transformation distinguishes Taylor powers modulo three and the second modulo six. For example,
\bea
f_+^{(8)}\sim t^2~,
 \hspace{0.4 in} f_-^{(8)}\sim t^5~.
\eea
The two powers have the same eigenvalue under $g_3^2$ but opposite eigenvalues under $g_3$. Fricke parity therefore removes the lower-order contribution and enforces an extra three powers of the displacement. This fact will supply the fifth-order entries needed in the quark construction below.

\subsection{Normalized modular form basis}

To evaluate the couplings away from the fixed point, we now choose a complete basis and fix its normalization independently of the vacuum. Using the Eisenstein combinations introduced above, take at level three
\bea\label{eq:ABCgenerators}
A\propto E_2(\tau)-3E_2(3\tau)~,\hspace{0.3 in}
 B\propto E_4(\tau)-9E_4(3\tau)~,\hspace{0.3 in}
 C\propto E_6(\tau)-27E_6(3\tau)~.
\eea
Their weights are $2,4,6$, and all three are Fricke odd. A complete basis is then
\bea\label{eq:levelthreebasis}
M_k^\epsilon(3)=\bigoplus_{\substack{2a+4b+6c=k\\a,c\geq0,\ b=0,1\\(-1)^{a+b+c}=\epsilon}}\mathbb C\,A^aB^bC^c~.
\eea
The restriction $b\leq1$ follows from the weight-eight relation $B^2=\kappa_4 A^4+\kappa_6 AC$, with constants determined by the normalization. Appendix \ref{app:modularspaces} proves completeness and gives the analogous level-two basis. In particular,
\bea\label{eq:emptylowweightspaces}
M_0^+(N)=\mathbb C~,\qquad M_0^-(N)=0~,\qquad M_2^+(N)=0~
\eea
for both $N=2,3$. The level-three generators have orders
\bea\label{eq:ABCorders}
\operatorname{ord}_{t_3=0}(A,B,C)=(2,1,0)~.
\eea

We now write $s_*=2\Im\tau_{3*}=1/\sqrt3$ and $(n_2,n_4,n_6)=(2,1,0)$, and define
\bea
F_k(\tau)=E_k(\tau)-3^{k/2}E_k(3\tau)~,\hspace{0.3 in}
 \ell_k={1\over n_k!}\left.{d^{n_k}\over dt^{n_k}}F_k(\tau(t))\right|_{t=0}~.
\eea
Our precise normalization convention is then to take
\bea
\label{eq:fixedpointnormalization}
A={F_2\over s_*\ell_2}~,\hspace{0.4 in}
 B={F_4\over s_*^2\ell_4}~,\hspace{0.4 in}
 C={F_6\over s_*^3\ell_6}~.
\eea
For a normalized generator $f$ of weight $k$ and order $n$, this gives $s_*^{k/2}\widehat f=t^n[1+O(t^6)]$, with canonical value
\bea\label{eq:normalizedcanonicalexpansion}
(2\Im\tau)^{k/2}f(\tau)=e^{ik\arg(1-t)}(1-|t|^2)^{k/2}t^n[1+O(t^6)]~.
\eea
The phase in this equation factorizes into field phases in each superpotential interaction and can be removed at a specified vacuum. The motivation for this normalization is that the leading term in the $t$ expansion has coefficient one. Appendix \ref{app:normalization} gives further details.

\subsection{A restriction at level two}
\label{sec:leveltwo}

The order-four and order-six fixed points provide different ranges of symmetry-enforced suppressions. At level two, every nonzero generic entry has order at most three by (\ref{eq:ordertables}). Consider a $3\times3$ Yukawa matrix whose allowed entries have their own real coefficients and whose determinant is generically nonzero. Each nonzero term in the determinant contains one entry from each row and column, so it begins at order at most nine. The leading terms cannot cancel for all choices of these coefficients. Therefore
\bea\label{eq:leveltwodeterminantbound}
\operatorname{ord}_{t_2=0}\det Y\leq9
\eea
on a generic set of coefficients. Exact zeros can remove determinant terms, but every remaining term still obeys this bound.

The product of the singular values equals $|\det Y|$. An unsuppressed largest singular value and phenomenologically desirable powers $(8,4,0)$ or $(7,3,0)$ would require determinant orders twelve or ten, respectively, and are therefore excluded for generic coefficients in this class of level-two models. Of course, this does not exclude all level-two models: the argument does not apply to tuned cancellations, correlations between entry coefficients, additional vanishing factors imposed on the coefficient spaces, or more general matter representations. However, it does suggest level three as a more natural starting point for comparison with phenomenology. Indeed, at level three a generic allowed entry can instead begin at fifth order, and the assignments below realize the desired quark hierarchy without any additional cancellations. For this reason, the concrete model studied in the following section will be for level three.

\subsection{Relation to non-invertible symmetry}
\label{sec:doublecosets}

Let us note in passing that the Fricke construction also has a connection to coset non-invertible symmetry.\footnote{Discussions of coset non-invertible symmetry can be found in \cite{Hsin:2024aqb,Hsin:2025ria,Putrov:2026csz}. For a more general review of non-invertible symmetries, see \cite{Kaidi:2026urc}. } To explain it, consider the finite group and subgroup
\bea\label{eq:finitecosetpair}
G_N=(\ZZ_N^T\times\ZZ_N^U)\rtimes\ZZ_2^W~,
 \qquad H=\langle U\rangle\simeq\ZZ_N~,
\eea
where $T$ and $U$ generate commuting cyclic groups of order $N$, and $W$ exchanges them: $WTW^{-1}=U$ and $WUW^{-1}=T$. Thus $H$ is not normal in $G_N$. In a theory with anomaly-free $G_N$ symmetry, gauging $H$ gives a standard example of a coset non-invertible symmetry \cite{Hsin:2024aqb,Hsin:2025ria}. The transformations generated by $T$ survive as invertible symmetries, while the transformation $W$, which exchanges the gauged subgroup with the other cyclic factor, becomes non-invertible after averaging over gauge transformations on both sides.

Indeed, a symmetry transformation can be represented by a topological defect across which the fields are transformed. After gauging $H$, independent gauge transformations on its two sides identify $g\in G_N$ with $h_Lgh_R$, for $h_L,h_R\in H$. The resulting operations are therefore organized by double cosets $HgH$. There are $N$ double cosets $HT^aH$, with $a=0,\ldots,N-1$, and one additional double coset $HWH$. Their action on $H$-invariant local operators can be written in the group algebra as
\bea\label{eq:cosetoperators}
e_H={1\over N}\sum_{b=0}^{N-1}U^b~,
 \qquad D_a=T^ae_H~,
 \qquad X=Ne_HWe_H~.
\eea
Here $e_H$ projects onto these operators and is the identity in the projected algebra. With this normalization, composition gives
\bea\label{eq:cosetoperatorfusion}
D_aD_b=D_{a+b}~,
 \qquad D_aX=XD_a=X~,
 \qquad X^2=N\sum_{a=0}^{N-1}D_a~,
\eea
where $a+b$ is understood modulo $N$. The last relation expresses the non-invertibility---composing $X$ with itself produces a sum of symmetry operations instead of the identity.\footnote{The full defect fusion algebra involves a condensation defect formed from the Wilson lines of the gauged $H$ symmetry; (\ref{eq:cosetoperatorfusion}) records only the induced action on local operators.}

For $N=2,3$, precisely this algebra appears in the modular realization. Let $\Gamma_N=\pm\Gamma(N)$, where $\Gamma(N)$ is the principal congruence subgroup. Then there is an isomorphism
\bea\label{eq:doublecosetconnection}
\mathbb C[\Gamma_N\backslash\Gamma_0^+(N)/\Gamma_N] \,\,\simeq\,\,\mathbb C[H\backslash G_N/H]~.
\eea
The modular double cosets represented by $T^a$ and $W_N$ correspond to $D_a$ and $X$, respectively. Appendix \ref{app:doublecosets} constructs the finite quotient and derives the isomorphism, including the normalization of the double-coset sums.

On the coefficient spaces used in this paper, every form is invariant under $T$, and the double-coset operators act as
\bea\label{eq:doublecosetcharacters}
D_af=f~,
 \qquad Xf=N(f|_kW_N)=\epsilon\, Nf~,
 \qquad f\in M_k^\epsilon(N)~.
\eea
Thus these spaces realize the two one-dimensional representations of the double-coset algebra in which every $D_a$ acts trivially. The double-coset viewpoint therefore embeds the two characters in a larger non-invertible algebra. The flavor interactions constructed in this work use only the ordinary sign action. More interesting non-invertible applications are left to future work. 

\section{Quark hierarchies, mixing, and CP violation}
\label{sec:quarks}
\label{sec:quarkhierarchies}

We now construct a level-three model using the ingredients in the previous section. The order-six fixed point allows the required separation of quark masses, while the same displacement controls the CKM angles.

\subsection{Matter assignment and local interactions}
\label{sec:N3model}

Choose the following characters and modular weights,
\bea\label{eq:quarkassignment}
\begin{array}{c|ccc}
 &1&2&3\\ \hline
 Q_i&(+,-4)&(-,4)&(+,14)\\
 U_i^c&(+,8)&(-,4)&(+,-10)\\
 D_i^c&(+,12)&(+,0)&(+,0)
\end{array}~,
\hspace{0.4 in}
H_u=(+,0)~,\quad H_d=(+,6)~.
\eea
In $W_Y=U_i^c\mathcal Y^u_{ij}Q_jH_u+D_i^c\mathcal Y^d_{ij}Q_jH_d$, rows label the right-handed conjugate fields and columns label the common left-handed space. The coefficient weights and Fricke signs are
\bea\label{eq:N3weightmatrices}\label{eq:quarksignmatrices}
K_u&=&\begin{pmatrix}4&12&22\\0&8&18\\-14&-6&4\end{pmatrix}~,
\hspace{0.34 in} K_d\,\,\,=\,\,\,\begin{pmatrix}14&22&32\\2&10&20\\2&10&20\end{pmatrix}~,\no\\
 \epsilon_u&=&\begin{pmatrix}+&-&+\\-&+&-\\+&-&+\end{pmatrix}~,
\hspace{0.7 in} \epsilon_d\,\,\,=\,\,\,\begin{pmatrix}+&-&+\\+&-&+\\+&-&+\end{pmatrix}~.
\eea
Let $f_\epsilon^{(k)}$ denote an arbitrary form in the entire weight-$k$ space of sign $\epsilon$. The holomorphic matrices are
\bea\label{eq:N3exactupYukawa}
{\mathcal Y_u\over y_0^u}=
\begin{pmatrix}
f_+^{(4)}&f_-^{(12)}&f_+^{(22)}\\
0&f_+^{(8)}&f_-^{(18)}\\
0&0&f_+^{(4)}
\end{pmatrix}~,
\hspace{0.4 in}
{\mathcal Y_d\over y_0^d}=
\begin{pmatrix}
f_+^{(14)}&f_-^{(22)}&f_+^{(32)}\\
0&f_-^{(10)}&f_+^{(20)}\\
0&f_-^{(10)}&f_+^{(20)}
\end{pmatrix}~.
\eea
Each occurrence of $f_\epsilon^{(k)}$ is a distinct function---for example, $f_+^{(4)}$ in the 11 and 33 entries of $\cY_u$ represent different weight-four modular forms. The up $21$ entry vanishes because there is no odd weight-zero form, while the up $31,32$ entries require negative weights. The down $21,31$ entries require even weight-two forms, which are absent. These are exact zeros of the holomorphic superpotential.

Using the normalized generators $A,B,C$ of weights $2,4,6$ and orders $2,1,0$, respectively, the least-vanishing form in each entry gives
\bea\label{eq:quarkleadingforms}
{\mathcal Y_u^{\rm min}\over y_0^u}=
\begin{pmatrix}
a_uA^2&b_uABC&c_uBC^3\\
0&d_uAC&e_uC^3\\
0&0&f_uA^2
\end{pmatrix}~, \hspace{0.2 in}
{\mathcal Y_d^{\rm min}\over y_0^d}=
\begin{pmatrix}
a_dA^2BC&b_dA^2C^3&c_dAC^5\\
0&d_dA^2C&e_dAC^3\\
0&f_dA^2C&g_dAC^3
\end{pmatrix}~.
\eea
The superscript distinguishes these representatives from the complete matrices used below. Higher basis elements in a given entry begin six or more powers later and do not change the generic leading orders.

\subsection{Mass and CKM hierarchies}
\label{sec:phenomenology}

With the minimal diagonal K\"ahler metric, canonical normalization introduces only smooth nonzero factors. The leading matrices therefore take the form
\bea\label{eq:quarktextures}
{Y_u\over y_0^u}\simeq
\begin{pmatrix}
a_ut^4&b_ut^3&c_ut\\0&d_ut^2&e_u\\0&0&f_ut^4
\end{pmatrix}~,
\hspace{0.4 in}
{Y_d\over y_0^d}\simeq
\begin{pmatrix}
a_dt^5&b_dt^4&c_dt^2\\0&d_dt^4&e_dt^2\\0&f_dt^4&g_dt^2
\end{pmatrix}~,
\eea
c.f. (\ref{eq:canonicalyukawa}).
The normalized local coefficients have been absorbed into the constants displayed here. The singular values are determined by the leading entry, minors, and determinant. In the up sector,
\bea\label{eq:N3updeterminant}
\det(Y_u/y_0^u)\sim a_ud_uf_u t^{10}~,
 \hspace{0.4 in}
 \det\begin{pmatrix}b_ut^3&c_ut\\d_ut^2&e_u\end{pmatrix}
 =(b_ue_u-c_ud_u)t^3~.
\eea
For $b_ue_u-c_ud_u\neq0$, this gives
\bea\label{eq:N3upmasses}
y_t\simeq y_0^u|e_u|~,
 \hspace{0.4 in} y_c\simeq y_0^u{|b_ue_u-c_ud_u|\over |e_u|}|t|^3~,
 \hspace{0.4 in} y_u\simeq y_0^u{|a_ud_uf_u|\over |b_ue_u-c_ud_u|}|t|^7~.
\eea
Note that the lightest mass is of order $|t|^7$, even though there is no order-seven term in (\ref{eq:quarktextures}) itself.

For the down sector, define
\bea\label{eq:N3downdeterminant}
\Delta_d&=&d_dg_d-e_df_d~,
\no\\
 M_d^2&=&|c_d|^2+|e_d|^2+|g_d|^2~,
 \no\\
 P_d^2&=&|b_de_d-c_dd_d|^2+|b_dg_d-c_df_d|^2+|\Delta_d|^2~.
\eea
The determinant is $a_d\Delta_d t^{11}$ at leading order, and $P_d$ is the norm of the order-six minors. Consequently,
\bea\label{eq:N3downmasses}
y_b\simeq y_0^d M_d|t|^2~,
 \hspace{0.4 in} y_s\simeq y_0^d{P_d\over M_d}|t|^4~,
 \hspace{0.4 in} y_d\simeq y_0^d{|a_d\Delta_d|\over P_d}|t|^5~.
\eea
For generic coefficients, the mass powers are therefore
\bea\label{eq:quarkmasspowers}
(y_u,y_c,y_t)\sim(|t|^7,|t|^3,1)~,
 \hspace{0.4 in} (y_d,y_s,y_b)\sim(|t|^5,|t|^4,|t|^2)~.
\eea

The CKM matrix compares the up- and down-quark mass bases in the common left-handed space spanned by $Q_i$. Since the columns of $Y_q$ label these fields, the left-handed rotations diagonalize $Y_q^\dagger Y_q$. In both sectors the heavy eigenvector approaches the third direction in the common $Q$ basis. Its first two components have orders $(t^5,t^2)$ in the up sector and $(t^3,t^2)$ in the down sector. After removing this direction, the rotation in each light two-dimensional subspace is of order $t$. The up and down coefficients are unrelated, so their rotations generically differ at these same orders and we have
\bea\label{eq:N3CKMpowers}
|V_{\rm CKM}|=|U_{uL}^\dagger U_{dL}|
 \sim\begin{pmatrix}
 1&|t|&|t|^3\\|t|&1&|t|^2\\|t|^3&|t|^2&1
 \end{pmatrix}~.
\eea
In particular, the product of a light rotation and a heavy $23$ rotation contributes at order $t^3$ to $V_{ub}$. We thus see that the same displacement supplies both the quark-mass and CKM hierarchies.

Note that the Fricke signs are essential to the above discussion. For example, if their constraints are removed while the weights and $T$ invariance are retained, the generic up matrix instead has orders
\bea
{Y_u\over y_0^u}\sim
\begin{pmatrix}t&1&t\\1&t^2&1\\0&0&t\end{pmatrix}~,
\eea
with singular-value powers $(|t|,1,1)$. The distinction between the two ordinary characters therefore protects the successful hierarchy.

Though we have assumed the diagonal K\"ahler metric above, any smooth positive-definite matter metrics whose eigenvalues remain bounded away from zero and infinity preserve the singular-value powers. Indeed, to determine their allowed flavor mixing, consider $g=TW_3$, which fixes $\tau_*={1\over2}+{i\over2\sqrt3}$. At this point $j_g(\tau_*)=\sqrt3\tau_*=e^{i\pi/6}$. Writing $\epsilon_{Q_i}=(-1)^{p_{Q_i}}$, the character and automorphy factor act together as
\bea
Q_i\longmapsto(-1)^{p_{Q_i}}e^{-i\pi k_{Q_i}/6}Q_i
 =e^{2\pi i q_i/6}Q_i~,
 \hspace{0.3 in}q_i=-{k_{Q_i}\over2}+3p_{Q_i}\pmod6~.
\eea
The assignment (\ref{eq:quarkassignment}) therefore gives $(q_1,q_2,q_3)=(2,1,5)$. Away from the fixed point, the local fields $\widetilde Q_i=(1-t)^{k_{Q_i}}Q_i$ obey the same constant phase transformation under $t\mapsto e^{-i\pi/3}t$. Writing their kinetic term as $\widetilde Q^\dagger Z_Q(t,\bar t)\widetilde Q$, invariance requires a Taylor monomial $t^a\bar t^b$ in $(Z_Q)_{ij}$ to satisfy $a-b\equiv q_j-q_i\pmod6$. Thus the first allowed terms in positions $12$, $13$, and $23$ are proportional to $\bar t$, to $t^3$ or $\bar t^3$, and to $\bar t^2$, respectively. Their magnitudes have orders $|t|$, $|t|^3$, and $|t|^2$. Regular canonical normalization consequently preserves the generic CKM powers, although it changes their coefficients. The exact zeros in (\ref{eq:N3exactupYukawa}) refer to the holomorphic basis and need not remain zeros after canonical normalization.

\subsection{Spontaneous CP violation}
\label{sec:spontaneousCP}

The same matter assignment admits a CP-invariant theory with real superpotential coefficients. Following the modular CP construction of \cite{Novichkov:2019sqv}, we impose
\bea\label{eq:CPtransformation}
\tau\longmapsto-\bar\tau~,\hspace{0.4 in}
 \Phi(x)\longmapsto\Phi^*(x_P)~.
\eea
Here $x_P=(x^0,-\mathbf x)$ is the parity-reflected spacetime point, and Lorentz and superspace transformations are implicit. On the modular group, CP induces the automorphism $T\mapsto T^{-1}$ and $W_3\mapsto W_3$ projectively. The latter equality can differ by $-\mathds1$ for matrix representatives, which acts trivially at the even weights used here. Both characters $+$ and $-$ are real and are preserved by this automorphism, so no enlargement of the matter representations is required.

The Fourier coefficients of $F_k(\tau)=E_k(\tau)-3^{k/2}E_k(3\tau)$ are real. At $\tau_*={1\over2}+{i\over2\sqrt3}$, the Fourier parameter is
\bea
q_*=e^{2\pi i\tau_*}=-e^{-\pi/\sqrt3}~,
\eea
which is real and negative. Hence $F_k(\tau_*)$ and $F_k''(\tau_*)$ are real, while $F_k'(\tau_*)$ is purely imaginary: each derivative multiplies a Fourier term by $2\pi i n$. Since $F_2$ has a double zero at $\tau_*$ and $\left.d\tau/dt\right|_{t=0}=is_*$, the normalization constants in (\ref{eq:fixedpointnormalization}) are
\bea
\ell_2=-{s_*^2\over2}F_2''(\tau_*)~,
 \hspace{0.3 in}\ell_4=is_*F_4'(\tau_*)~,
 \hspace{0.3 in}\ell_6=F_6(\tau_*)~,
\eea
and are all real. Thus the fixed normalization of $A$, $B$, and $C$ introduces no complex phases, and their real Fourier coefficients imply
\bea\label{eq:CPgenerators}
(A,B,C)(-\bar\tau)=\overline{(A,B,C)(\tau)}~.
\eea
Taking every coefficient in the complete monomial bases and both overall Yukawa factors to be real therefore ensures CP invariance of the flavor superpotential. The diagonal matter metric is compatible with the same transformation, so a complex vev for $\tau$ can break CP spontaneously.

The leading terms in the local Taylor expansions of the modular coefficients do not produce a CKM phase. To see this, write $t=r e^{i\varphi}$ and remove the field-dependent automorphy phases in (\ref{eq:normalizedcanonicalexpansion}). The leading matrices (\ref{eq:quarktextures}) factor as
\bea\label{eq:CPleadingfactorization}
Y_u^{(0)}=D_u\,Y_{u,\mathbb R}^{(0)}P~,
 \hspace{0.4 in}
 Y_d^{(0)}=D_d\,Y_{d,\mathbb R}^{(0)}P~,
 \hspace{0.4 in}
 P=\operatorname{diag}(e^{3i\varphi},e^{2i\varphi},1)~,
\eea
with $D_u=\operatorname{diag}(e^{i\varphi},1,e^{4i\varphi})$, $D_d=e^{2i\varphi}\,\mathds1$, and real matrices $Y_{q,\mathbb R}^{(0)}$. Right-handed field rephasings remove $D_u,D_d$, and the common left-handed rephasing removes $P$. Thus a calculation retaining only these leading powers would miss the source of CP violation.

On the other hand, the full modular forms have phases that need not obey this leading relation. A useful invariant is
\bea\label{eq:CPgeneratorratio}
\eta_{\rm CP}(\tau)=\arg{B(\tau)^2\over A(\tau)C(\tau)}~.
\eea
The ratio has weight zero and positive Fricke parity, so it is modular invariant wherever defined, and CP conjugates it. Its role in the quark matrices is particularly direct. Put $\alpha=\arg A$, $\beta=\arg B$, and $\gamma=\arg C$. With real coefficients, each of the representative matrices (\ref{eq:quarkleadingforms}) can separately be made real by row and column rephasings. Their column phase matrices may be chosen as
\bea\label{eq:CPcolumnphases}
P_u&=&\operatorname{diag}\bigl(e^{i(2\alpha-\beta-3\gamma)}~,
                         e^{i(\alpha-2\gamma)},1\bigr)~,\no\\
 P_d&=&\operatorname{diag}\bigl(e^{i(\alpha+\beta-4\gamma)}~,
                         e^{i(\alpha-2\gamma)},1\bigr)~.
\eea
Their mismatch is $P_dP_u^\dagger=\operatorname{diag}(e^{i\eta_{\rm CP}},1,1)$, since $\eta_{\rm CP}=2\beta-\alpha-\gamma$. Accordingly, these representatives give $V_{\rm CKM}=O_u^T\operatorname{diag}(e^{-i\eta_{\rm CP}},1,1)O_d$, up to unphysical phases, with real orthogonal $O_u,O_d$. For generic coefficients the relative phase cannot be removed from the CKM matrix. 

The size of the effect follows from the exact generator relation (\ref{eq:normalizedgeneratorrelation}). Since $B^2$ and $AC$ have the same normalized leading term, $8\ell_2\ell_6/\ell_4^2=1$, and hence
\bea\label{eq:CPexactrelation}
B^2=AC-\kappa A^4~,\hspace{0.4 in}
 {B^2\over AC}=1-\kappa{A^3\over C}~,\hspace{0.4 in}
 \kappa={22\ell_2^4\over \ell_4^2}\simeq406.5~.
\eea
Near the fixed point, $A^3/C=t^6[1+O(t^6)]$. Although the phase therefore vanishes at leading order in the displacement, the fixed coefficient multiplying this correction is appreciable in our normalization. At the numerical fit below, the exact values give
\bea\label{eq:CPfitratio}
\left.{B^2\over AC}\right|_{\tau_0}
 =0.8163-0.2048\,i~,
 \hspace{0.4 in} \eta_{\rm CP}(\tau_0)=-0.2458~.
\eea
This supplies a substantial relative phase with entirely real couplings. Let us emphasize that the numerical fit evaluates the full modular forms and canonical factors in the fixed normalization of Appendix \ref{app:normalization}.

\subsection{A fit with real coefficients}
\label{sec:quarkfit}
\label{sec:jointnumericalfit}

The symmetry and matter assignments determine the generic hierarchy powers derived above, while the precise masses and mixing parameters depend on the remaining coefficients, which we now fit to the reference values. To construct a fit, include every basis form in (\ref{eq:N3exactupYukawa}). In each entry, vary the real coefficient of the least-vanishing form and set the coefficients of the higher basis forms to $+1$. The complete bases and remaining thirteen real coefficients are given in Appendix \ref{app:quarkfit}. 

In particular, take
\bea\label{eq:numericaltvalue}
\label{eq:quarkoverallscales}
t_0&=&0.300503+0.0567136i~,\no\\
 \tau_0&=&0.433517+0.531314i~,
\eea
with the thirteen leading coefficients given in Appendix \ref{app:quarkfit}. Together with the twenty-five higher coefficients and the two overall factors, all quark coefficients are real and take values in $[1/4,4]$, while the displacement has magnitude $|t_0|=0.305808$.  

\begin{figure}[p]
\centering
\includegraphics[width=0.97\textwidth]{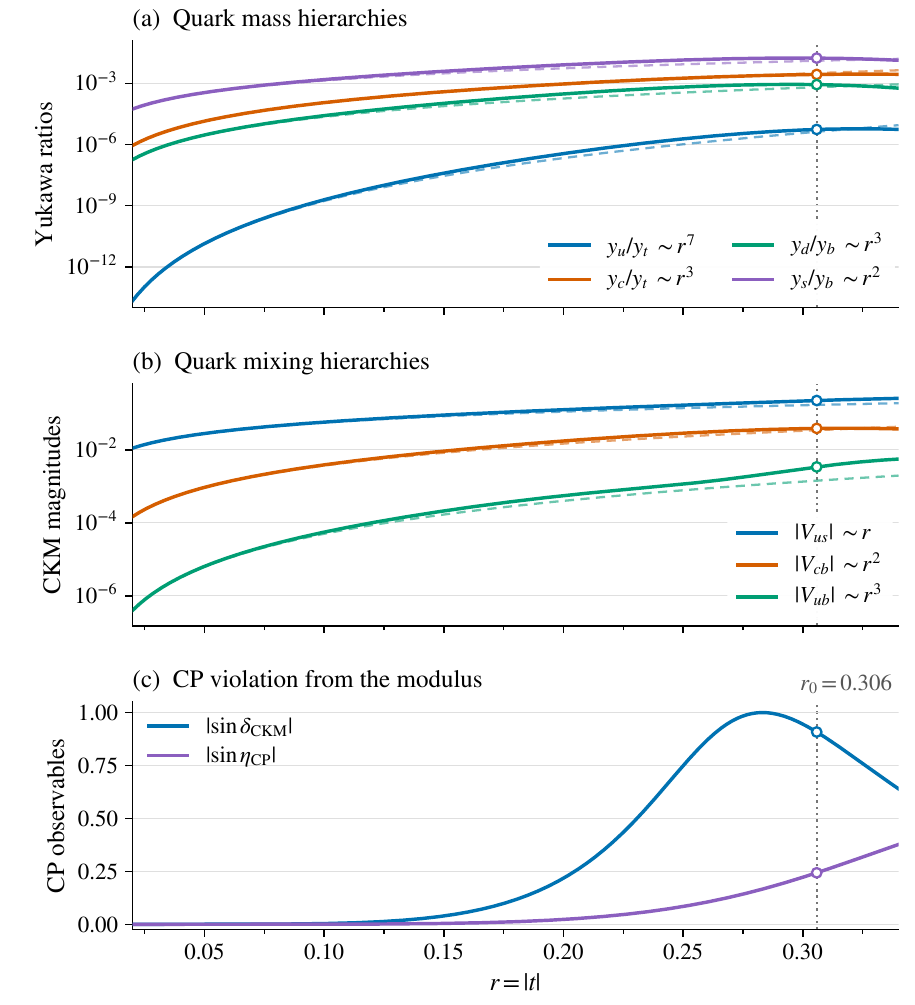}
\caption{Quark hierarchies and spontaneous CP violation along $t=r e^{i\arg t_0}$, with all coefficients fixed at the fit values. Solid curves use the complete modular forms and diagonal canonical normalization. Dashed curves in panels (a) and (b) show the indicated asymptotic powers, matched to the solid curves at $r=0.02$. Panel (c) shows $|\sin\delta_{\rm CKM}|$ and $|\sin\eta_{\rm CP}|$; both vanish as $r^6$ near the fixed point but become appreciable at finite displacement. The vertical line and open circles mark the fitted vacuum $r_0=0.306$.}
\label{fig:quarkhierarchies}
\end{figure}

For comparison with quark data, we adopt the $\tan\beta=10$ high-scale values of Table 6 in \cite{Antusch:2025fpm}, with $M_{\rm SUSY}=3\,{\rm TeV}$, the vanishing finite supersymmetric threshold corrections used there, and $M_{\rm GUT}=2\times10^{16}\,{\rm GeV}$. Their SM-equivalent Yukawas are divided by $\sin\beta$ in the up sector and $\cos\beta$ in the down sector. Evaluating the exact modular forms with these inputs gives
\bea\label{eq:quarkfitobservables}
\begin{array}{c|c@{\hspace{0.4 in}}c|c}
\text{quantity}&\text{value}&\text{quantity}&\text{value}\\ \hline
 y_u&2.774\times10^{-6}&y_d&4.924\times10^{-5}\\
 y_c&1.397\times10^{-3}&y_s&9.849\times10^{-4}\\
 y_t&0.4972&y_b&0.05527\\
 \theta_{12}&0.2270&\theta_{23}&0.03874\\
 \theta_{13}&0.003410&\delta_{\rm CKM}&1.139
\end{array}~,
\eea
where angles are in radians and the phase uses the standard CKM parametrization. The resulting mixing matrix and Jarlskog invariant are
\bea\label{eq:numericalCKMmatrix}
|V_{\rm CKM}|=\begin{pmatrix}
 0.9743&0.2251&0.003410\\
 0.2249&0.9736&0.03873\\
 0.007923&0.03806&0.9992
 \end{pmatrix}~,
 \hspace{0.4 in} J_{\rm CKM}=2.628\times10^{-5}~.
\eea
Figure \ref{fig:quarkhierarchies} shows the approach to the asymptotic hierarchies along $\arg t=\arg t_0$, with all coefficients held fixed. The same figure displays the CKM phase generated by the exact modular forms. The leading mass-generating combinations have
\bea
{|b_ue_u-c_ud_u|\over |b_ue_u|+|c_ud_u|}=0.3635~,
 \hspace{0.4 in}
 {|d_dg_d-e_df_d|\over |d_dg_d|+|e_df_d|}=0.4456~,
\eea
so the quark fit does not rely on severe cancellations in these combinations. As a check on the origin of CP violation, replacing $t_0$ by $\bar t_0$ leaves all masses and CKM magnitudes unchanged and reverses the sign of $J_{\rm CKM}$. Here $t\mapsto\bar t$ corresponds to $\tau\mapsto1-\bar\tau$, namely CP followed by $T$.\footnote{ Let us mention that the strong-CP problem depends on the gluon gauge kinetic function and threshold corrections, and is separate from the CKM phase studied here \cite{Feruglio:2023uof}.}

\section{Lepton masses and mixing}
\label{sec:leptons}
\label{sec:Majorana}

The same Fricke symmetry accommodates charged-lepton hierarchies and large lepton mixing. We supplement the MSSM by three Standard Model singlets $N_i^c$ and use a type-I seesaw. The singlets are odd under matter parity. All lepton fields are given positive Fricke parity, with weight vectors listed in generation order,
\bea\label{eq:ordinaryleptonassignment}
k_L=(6,6,6)~,\qquad k_{E^c}=(2,4,8)~,\qquad k_{N^c}=(12,12,10)~.
\eea
The Higgs assignments remain $(+,0)$ for $H_u$ and $(+,6)$ for $H_d$. The lepton superpotential is
\bea\label{eq:ordinaryleptonsuperpotential}
W_\ell=E_i^c(\mathcal Y_e)_{ij}L_jH_d
 +N_i^c(\mathcal Y_\nu)_{ij}L_jH_u
 +{1\over 2}N_i^c(\mathcal M_N)_{ij}N_j^c~.
\eea
The coefficient functions then have positive Fricke parity and weights
\bea\label{eq:ordinaryleptonweights}
K_e=\begin{pmatrix}14&14&14\\16&16&16\\20&20&20\end{pmatrix}~,\hspace{0.4 in}
 K_\nu=\begin{pmatrix}18&18&18\\18&18&18\\16&16&16\end{pmatrix}~,\hspace{0.4 in}
 K_N=\begin{pmatrix}24&24&22\\24&24&22\\22&22&20\end{pmatrix}~.
\eea
Every entry admits an independent linear combination of the complete modular-form basis at its indicated weight.

\subsection{Hierarchy and the seesaw}
\label{sec:leptonhierarchies}

For the assignments above, the least-suppressed basis elements give
\bea\label{eq:ordinarychargedtexture}\label{eq:ordinarydiractexture}\label{eq:ordinarymajoranatexture}
\mathcal Y_e&=&\operatorname{diag}(A^2BC,A^2C^2, AC^3)c_e+\cdots~,
 \no\\
 \mathcal Y_\nu&=&\operatorname{diag}(ABC^2,ABC^2,A^2C^2)c_\nu+\cdots~,
 \no\\
 \mathcal M_N&=&M_0\begin{pmatrix}
 m_{11}C^4&m_{12}C^4&m_{13}BC^3\\
 m_{12}C^4&m_{22}C^4&m_{23}BC^3\\
 m_{13}BC^3&m_{23}BC^3&m_{33}AC^3
 \end{pmatrix}+\cdots~.
\eea
Here $c_e$ and $c_\nu$ are general real matrices, $m_{ij}=m_{ji}\in\mathbb R$, and the ellipses denote the remaining allowed basis elements. Since $A\sim t^2$, $B\sim t$, and $C\sim1$, the charged-lepton singular values have the generic hierarchy
\bea\label{eq:ordinarychargedhierarchy}
y_e:y_\mu:y_\tau\sim |t|^5:|t|^4:|t|^2~.
\eea
For example, the determinant has order eleven, the least-suppressed $2\times2$ minors have order six, and the largest entry has order two. As before, smooth, nonsingular canonical normalization preserves these powers. Because the three $L_i$ have identical modular weights and Fricke parities, all three entries within each row of $Y_e$ have the same suppression. Their independent coefficients therefore allow order-one mixing among the left-handed charged leptons, despite the hierarchy of their masses.

The neutrino powers have a simple factorized form. Writing $D=\operatorname{diag}(1,1,t)$, the leading matrices can be expressed as
\bea
Y_\nu=t^3D\,\widehat Y_\nu~,\hspace{0.4 in}
 M_N=M_0D\,\widehat M_ND~,
\eea
where the hatted matrices have finite, generically nonsingular limits at $t=0$. The smooth factors from canonical normalization can be included in these matrices. Integrating out the three singlets gives
\bea\label{eq:ordinaryseesaw}
M_\nu=-v_u^2Y_\nu^TM_N^{-1}Y_\nu
 =-{v_u^2t^6\over M_0}\widehat Y_\nu^T\widehat M_N^{-1}\widehat Y_\nu~.
\eea
The additional suppression of the third Dirac row cancels against the corresponding Majorana factors. Thus the three light-neutrino masses have the same generic order $v_u^2|t|^6/M_0$, while the heavy masses have orders $M_0(1,1,|t|^2)$. 

The symmetry also permits a direct Weinberg operator, with a positive-parity weight-twelve coefficient. Its generic contribution has order $v_u^2/\Lambda$, since this coefficient space contains $C^2$. We assume that its ultraviolet scale is sufficiently large for the seesaw contribution to dominate. For example, $\Lambda\sim10^{18}\,{\rm GeV}$ gives $v_u^2/\Lambda\sim3\times10^{-5}\,{\rm eV}$, below the masses in the fit below.

\subsection{A numerical example}
\label{sec:completeleptons}

At the same vacuum (\ref{eq:numericaltvalue}) as the quark fit, real lepton coefficients reproduce the charged-lepton reference values of \cite{Antusch:2025fpm} and a normal-ordering neutrino spectrum with the representative oscillation parameters of NuFIT~6.1 \cite{Esteban:2024eli}.\footnote{We use the normal-ordering IC24 fit including SK atmospheric data from the November 2025 NuFIT 6.1 update; see the \href{https://www.nu-fit.org/sites/default/files/v61.tbl-parameters.pdf}{parameter table}.} We include every allowed basis monomial with a real coefficient of order one. The fit gives
\bea
(y_e,y_\mu,y_\tau)=(2.020\times10^{-5},\,0.004255,\,0.07259)~.
\eea
With $v_u\simeq174\,{\rm GeV}$ (using $\sin\beta\simeq1$ at $\tan\beta=10$) and $M_0=2.26867\times10^{12}\,{\rm GeV}$, the light-neutrino masses and mixing parameters are
\bea
(m_1,m_2,m_3)&=&(5.000,\,10.02,\,50.36)\,{\rm meV}~,\no\\
 (\Delta m_{21}^2,\Delta m_{31}^2)&=&(7.537\times10^{-5},\,2.511\times10^{-3})\,{\rm eV}^2~,\no\\
 (\sin^2\theta_{12},\sin^2\theta_{13},\sin^2\theta_{23})
 &=&(0.3088,\,0.02248,\,0.4700)~,\quad \delta_{\rm PMNS}=180^\circ~.
\eea
Here $U_{eL}$ diagonalizes $Y_e^\dagger Y_e$, $U_\nu$ Takagi-diagonalizes $M_\nu$, and $U_{\rm PMNS}=U_{eL}^\dagger U_\nu$. The physical heavy-neutrino masses are approximately
\bea
(M_1,M_2,M_3)=(2.424\times10^{11},\,2.172\times10^{12},\,2.275\times10^{12})\,{\rm GeV}~.
\eea
Appendix \ref{app:leptonfit} specifies all coefficients. Their magnitudes range from $0.25$ to $3.914$. The electron mass involves a cancellation in the leading charged-lepton determinant, whose magnitude is $2.2\%$ of the sum of the absolute values of its six permutation terms. Note that the neutrino quantities are evaluated at tree level, using the NuFIT values directly as targets without renormalization-group evolution.\footnote{
For renormalization-group evolution in seesaw models,
including the successive decoupling of heavy neutrinos
at their mass thresholds, see \cite{Antusch:2005gp}.}

\section*{Acknowledgements}

J.K. is supported by JSPS KAKENHI Grant No. 26K17152 and through the Inamori Frontier Program at Kyushu University. H.O. was supported by JSPS KAKENHI Grant Numbers JP25H01539 and JP26K07087. T.K. and K.K. are supported by K2-SPRING program Grant number JPMJSP2136. The numerical calculations were assisted by OpenAI's ChatGPT, which identified the vacuum (\ref{eq:numericaltvalue}) and coefficients in Appendix \ref{app:numericaldata}. 

\appendix

\section{Complete modular form spaces at levels two and three}
\label{app:modularspaces}

In this appendix, we establish the complete scalar coefficient bases used in the main text and show that their generic leading powers saturate (\ref{eq:fixedpointcongruence}). All weights are even, all forms have trivial character on $\Gamma_0(N)$, and regularity is imposed at every cusp. We use the dimension and valence formulas for congruence subgroups; standard constructions of these spaces are reviewed in \cite{DHoker:2024xor}. For the levels of interest,
\bea\label{eq:dimensionssmalllevels}
\dim M_k(\Gamma_0(2))=\left\lfloor{k\over 4}\right\rfloor+1~,
 \hspace{0.4 in}
 \dim M_k(\Gamma_0(3))=\left\lfloor{k\over 3}\right\rfloor+1
\eea
for nonnegative even $k$. The index of $\Gamma_0(N)$ in the full modular group is three at level two and four at level three. Thus the valence formula counts zeros, with elliptic multiplicities divided by their stabilizer orders, to total $k/4$ and $k/3$, respectively.

\subsection{Level two}

Define the Fricke-odd forms
\bea
P=E_2(\tau)-2E_2(2\tau)~,\hspace{0.4 in}
 Q=E_4(\tau)-4E_4(2\tau)
\eea
of weights two and four. The elliptic point $\tau_{2*}$ has stabilizer of order two in $\Gamma_0(2)$, while its stabilizer in the Fricke extension has order four. Equation (\ref{eq:fixedpointcongruence}) requires the orders of $P$ and $Q$ to be congruent to one and zero modulo four. The valence formula then implies
\bea
\operatorname{ord}_{\tau_{2*}}P=1~,\hspace{0.4 in}
 \operatorname{ord}_{\tau_{2*}}Q=0~.
\eea
Indeed, a larger allowed order would exceed the total valence in either case.

The monomials $P^aQ^b$ with $2a+4b=k$ have distinct orders $a$ and are therefore linearly independent. Their number is $\lfloor k/4\rfloor+1$, equal to the dimension in (\ref{eq:dimensionssmalllevels}). They form a complete basis, and resolving their Fricke parities gives
\bea\label{eq:leveltwocompletebasis}
M_k^\epsilon(2)=
 \bigoplus_{\substack{2a+4b=k\\a,b\geq0\\(-1)^{a+b}=\epsilon}}
 \mathbb C\,P^aQ^b~.
\eea
Writing $\epsilon=(-1)^p$, the order of every basis element obeys
\bea
a\equiv2p-{k\over 2}\pmod4~.
\eea
If an allowed monomial has $a\geq4$, replacing $(a,b)$ by $(a-4,b+2)$ preserves its weight and parity and lowers its order by four. Every nonempty space therefore contains the unique monomial whose order is the least nonnegative residue in this congruence class. A generic linear combination has that order, proving the level-two part of (\ref{eq:ordertables}) at all weights.

\subsection{Level three}

For clarity, begin with the unnormalized forms
\bea
F_2=E_2(\tau)-3E_2(3\tau)~,\quad
 F_4=E_4(\tau)-9E_4(3\tau)~,\quad
 F_6=E_6(\tau)-27E_6(3\tau)~.
\eea
They are Fricke odd. At $\tau_{3*}$, which has an order-three stabilizer in $\Gamma_0(3)$, their allowed orders are respectively two, one, and zero modulo six. The valence formula forces the first two orders to be exactly two and one. For $F_6$, a positive order would have to be six and would exhaust its entire valence. Since $F_2^3$ has exactly that divisor, this would force $F_6$ to be proportional to $F_2^3$. Their Fourier expansions,
\bea
F_2=-2-24q+O(q^2)~,\hspace{0.4 in}
 F_6=-26-504q+O(q^2)~,
\eea
exclude proportionality. Therefore $F_6(\tau_{3*})\neq0$, and the orders of the normalized generators $(A,B,C)$ are $(2,1,0)$.

Consider all monomials $A^aB^bC^c$ satisfying $2a+4b+6c=k$, with $b=0,1$. Their orders are $2a+b$, which are distinct---the parity of the order determines $b$, its value then determines $a$, and the weight determines $c$. Hence these forms are linearly independent. Their number is $\lfloor k/3\rfloor+1$, equal to the full dimension in (\ref{eq:dimensionssmalllevels}). They form a complete basis of $M_k(\Gamma_0(3))$. Sorting them by their parity $(-1)^{a+b+c}$ proves (\ref{eq:levelthreebasis}).

In particular, the even weight-eight space is spanned by $F_2^4$ and $F_2F_6$. Matching constant and first Fourier coefficients gives the exact relation
\bea\label{eq:rawgeneratorrelation}
F_4^2=8F_2F_6-22F_2^4~.
\eea
For the normalization (\ref{eq:fixedpointnormalization}), this becomes
\bea\label{eq:normalizedgeneratorrelation}
B^2=-{22\ell_2^4\over \ell_4^2}A^4
 +{8\ell_2\ell_6\over \ell_4^2}AC~.
\eea
Thus powers $B^b$ with $b\geq2$ introduce no additional coefficient functions.

Let $n=2a+b$ be the order of a basis element with parity $(-1)^p$. Its weight obeys
\bea
n+{k\over 2}=3(a+b+c)~,
 \hspace{0.4 in} n\equiv3p-{k\over 2}\pmod6~.
\eea
Whenever $n\geq6$ one has $a\geq3$, and replacing $(a,b,c)$ by $(a-3,b,c+1)$ preserves weight and parity while reducing $n$ by six. Every nonempty coefficient space therefore contains exactly one basis element with the least nonnegative allowed order. This proves the level-three part of (\ref{eq:ordertables}), including at arbitrarily high weight. At weight zero only the even constant exists; at weight two only the odd generator exists.

\subsection{Genericity assumption}

As we have just seen, in both bases the smallest-order monomial in a nonempty character space is unique. Its coefficient must be nonzero for the generic order to be attained. Setting it to zero selects a proper linear subspace and raises the order, sometimes to an exact zero. Our genericity assumption is that this coefficient is nonzero.

For the Yukawa matrices, each allowed entry has its own real coefficients, even when several entries use the same modular form. Whenever the allowed entries permit a nonzero determinant or minor, its leading terms can cancel only for special choices of these coefficients. We assume that no such cancellation occurs when deriving the mass and mixing powers.

\section{Normalization of modular coefficients}
\label{app:normalization}

\subsection{The fixed-point convention}

For an interaction $W=\lambda F(\tau)\cO(\Phi)$, multiplying $F$ by a nonzero constant and dividing $\lambda$ by the same constant leaves the physics unchanged. A statement that $\lambda$ is of order one therefore requires a normalization convention \cite{Petcov:2023fwh}. We fix this convention at the symmetry point, before choosing a vacuum or fitting the coefficients. The prescription gives the coefficients a direct interpretation as local couplings to the modulus in fields canonical at that point.

For the level-three model, write $\tau_*=\tau_{3*}$ and define
\bea
t={\tau-\tau_*\over\tau-\bar\tau_*}~,
 \qquad s_*=2\Im\tau_*={1\over\sqrt3}~.
\eea
For a nonzero holomorphic form $F$ of weight $k$ and leading order $n$, define
\bea
\ell_F=\lim_{t\to0}{F(\tau(t))\over t^n}~,
 \qquad f(\tau)={F(\tau)\over s_*^{k/2}\ell_F}~.
\eea
Dividing by $\ell_F$ fixes the first nonzero Taylor coefficient, while the factor $s_*^{k/2}$ accounts for the matter normalization at $\tau_*$. Applying this prescription to the three generators gives (\ref{eq:fixedpointnormalization}). Their products then fix the entire monomial basis (\ref{eq:levelthreebasis}); no further vacuum-dependent rescaling is made. In particular, for any of these normalized monomials $f$,
\bea\label{eq:appnormalizedform}
f(\tau(t))=s_*^{-k/2}(1-t)^k t^n\cG_f(t)~,
 \qquad \cG_f(t)=1+O(t^6)~.
\eea
The factor $(1-t)^k$ restores the varying automorphy factor removed in (\ref{eq:covariantdiscframe}). 

\subsection{Physical fields and local interactions}

The relation between the upper-half-plane and disc coordinates is
\bea
2\Im\tau=s_*{1-|t|^2\over|1-t|^2}~.
\eea
Consequently the holomorphic matter redefinition
\bea\label{eq:applocalmatter}
\Psi_\Phi=s_*^{-k_\Phi/2}(1-t)^{k_\Phi}\Phi
\eea
puts the diagonal matter terms in (\ref{eq:kahlerpotential}) in the form $\Psi_\Phi^\dagger e^{2V}\Psi_\Phi/(1-|t|^2)^{k_\Phi}$. These variables are thus canonical at $t=0$. At a finite vacuum $t_0$, they must be rescaled further by $(1-|t_0|^2)^{-k_\Phi/2}$ to become canonical.

For a Yukawa interaction, $\cO(\Phi)$ is a product of three chiral fields and $\lambda$ is dimensionless. Since the coefficient weight $k$ equals the sum of their weights, the factors of $s_*$ and $(1-t)$ cancel between (\ref{eq:appnormalizedform}) and (\ref{eq:applocalmatter}), giving
\bea\label{eq:localnormalizedinteraction}
W=\lambda f(\tau)\cO(\Phi)=\lambda\,t^n\cG_f(t)\cO(\Psi)~.
\eea
Because $\cG_f(0)=1$, $\lambda$ is the coefficient of the leading power of $t$. To express this interaction in physical units, recall that at vanishing matter fields the modulus metric is
\bea
K_{t\bar t}={\Lambda_K^2\over(1-|t|^2)^2}~,
\qquad \varphi=\Lambda_Kt~,
\eea
where $\varphi$ has mass dimension one and is canonically normalized to leading order near $t=0$. The interaction therefore becomes
\bea
\label{eq:appmoduluscoupling}
W=\lambda\left({\varphi\over\Lambda_K}\right)^n
\cG_f\!\left({\varphi\over\Lambda_K}\right)\cO(\Psi)~.
\eea

At a vacuum $t=t_0$, substituting the modulus vev and canonically normalizing the matter fields gives
\bea
y(t_0)=\lambda(1-|t_0|^2)^{k/2}t_0^n\cG_f(t_0)~.
\eea
Thus an order-one interaction coefficient $\lambda$ can generate a small Yukawa coupling through the factor $t_0^n$. This expression differs from (\ref{eq:normalizedcanonicalexpansion}) only by field rephasings. The same argument applies to Majorana masses and higher-dimensional operators, with their dimensionful overall factors retained separately.

One caveat is that the scale $\Lambda_K$ is fixed by the modulus kinetic term and need not equal the scale suppressing its interactions. If the leading interaction is instead written as $c(\varphi/M)^n\cO(\Psi)$, comparison with (\ref{eq:appmoduluscoupling}) gives
\bea\label{eq:appinteractionnormalization}
\lambda=c\left({\Lambda_K\over M}\right)^n~.
\eea
An order-one value of $\lambda$ is therefore motivated if $c$ is of order one and $M$ is sufficiently close to $\Lambda_K$. This is an additional assumption about the underlying interactions.

\section{Numerical inputs}
\label{app:numericaldata}

This appendix collects the complete modular bases and coefficients used for the quark and lepton fits at the vacuum (\ref{eq:numericaltvalue}).

\subsection{Quark sector}
\label{app:quarkfit}

The complete bases used in Section \ref{sec:quarkfit} are listed below, ordered by increasing vanishing order at $\tau_*$,
\bea\label{eq:numericalfullbases}
\begin{array}{c|l}
 (k,\epsilon)&\text{ordered basis}\\ \hline
 (4,+)&(A^2)\\
 (8,+)&(AC,A^4)\\
 (10,-)&(A^2C,A^5)\\
 (12,-)&(ABC,A^4B)\\
 (14,+)&(A^2BC,A^5B)\\
 (18,-)&(C^3,A^3C^2,A^6C,A^9)\\
 (20,+)&(AC^3,A^4C^2,A^7C,A^{10})\\
 (22,+)&(BC^3,A^3BC^2,A^6BC,A^9B)\\
 (22,-)&(A^2C^3,A^5C^2,A^8C,A^{11})\\
 (32,+)&(AC^5,A^4C^4,A^7C^3,A^{10}C^2,A^{13}C,A^{16})
\end{array}~.
\eea
These are the monomials $A^aB^bC^c$ with $b=0,1$, $2a+4b+6c=k$, and $(-1)^{a+b+c}=\epsilon$. For a basis $(M_0,M_1,\ldots)$, define
\bea
\mathcal P_{k,\epsilon}(c)=cM_0+\sum_{j\geq1}M_j~.
\eea
The exact matrices of the fit are then
\bea\label{eq:numericaljointtextures}
{\mathcal Y_u\over y_0^u}&=&
 \begin{pmatrix}
 \mathcal P_{4,+}(a_u)&\mathcal P_{12,-}(b_u)&\mathcal P_{22,+}(c_u)\\
 0&\mathcal P_{8,+}(d_u)&\mathcal P_{18,-}(e_u)\\
 0&0&\mathcal P_{4,+}(f_u)
 \end{pmatrix}~,\no\\
 {\mathcal Y_d\over y_0^d}&=&
 \begin{pmatrix}
 \mathcal P_{14,+}(a_d)&\mathcal P_{22,-}(b_d)&\mathcal P_{32,+}(c_d)\\
 0&\mathcal P_{10,-}(d_d)&\mathcal P_{20,+}(e_d)\\
 0&\mathcal P_{10,-}(f_d)&\mathcal P_{20,+}(g_d)
 \end{pmatrix}~,
\eea
where the real leading coefficients are
\bea\label{eq:numericalquarkcoefficients}
\begin{array}{c|r@{\hspace{0.4 in}}c|r}
 a_u& 0.303498&a_d&-1.75703\\
 b_u& 0.25&b_d& 0.967454\\
 c_u&-3.77374&c_d& 3.84841\\
 d_u&-0.355063&d_d&-0.297808\\
 e_u& 2.50191&e_d& 0.250031\\
 f_u& 0.284097&f_d&-0.272789\\
 &&g_d&0.597192
\end{array}~.
\eea
The overall factors are given by 
\bea
y_0^u&=&0.250922~,\hspace{0.4 in} y_0^d=0.273954~.
\eea

\subsection{Lepton sector}
\label{app:leptonfit}

We next specify the real coefficients of the fit in Section \ref{sec:completeleptons}.
As before, for each weight, multiply the least-vanishing monomial by a real coefficient $x$ and set the coefficients of all remaining monomials to $+1$. The complete positive-parity spaces used here then give
\bea
\mathcal P_{14}(x)&=&xA^2BC+A^5B~,\no\\
 \mathcal P_{16}(x)&=&xA^2C^2+A^5C+A^8~,\no\\
 \mathcal P_{18}(x)&=&xABC^2+A^4BC+A^7B~,\no\\
 \mathcal P_{20}(x)&=&xAC^3+A^4C^2+A^7C+A^{10}~,\no\\
 \mathcal P_{22}(x)&=&xBC^3+A^3BC^2+A^6BC+A^9B~,\no\\
 \mathcal P_{24}(x)&=&xC^4+A^3C^3+A^6C^2+A^9C+A^{12}~.
\eea
Evaluating these polynomials at $\tau_0$ and applying (\ref{eq:canonicalyukawa}), the canonically normalized matrices are specified entry by entry in terms of the weight matrices (\ref{eq:ordinaryleptonweights}) as
\bea
(Y_e)_{ij}&=&(2\Im\tau_0)^{(K_e)_{ij}/2}\mathcal P_{(K_e)_{ij}}((c_e)_{ij})~,\no\\
 (Y_\nu)_{ij}&=&(2\Im\tau_0)^{(K_\nu)_{ij}/2}\mathcal P_{(K_\nu)_{ij}}((c_\nu)_{ij})~,\no\\
 (M_N)_{ij}&=&M_0(2\Im\tau_0)^{(K_N)_{ij}/2}\mathcal P_{(K_N)_{ij}}(m_{ij})~.
\eea
The real leading coefficients are then
\bea
c_e&=&\begin{pmatrix}
 0.25&0.25&0.25\\
 0.25&0.816591&0.25\\
 0.760408&0.25&0.832147
 \end{pmatrix}~,\no\\
 c_\nu&=&\begin{pmatrix}
 -0.630006&2.20686&1.86431\\
 0.250450&-1.81835&-0.500259\\
 1.55845&3.91349&2.57768
 \end{pmatrix}~,\no\\
 m&=&\begin{pmatrix}
 1&1&1\\
 1&-1&1\\
 1&1&-1
 \end{pmatrix}~.
\eea
The mass scales are
\bea
M_0=2.26867\times10^{12}\,{\rm GeV}~,\hspace{0.4 in}
 v_u=174\,{\rm GeV}~.
\eea

\section{Mixed modular-gauge anomalies}
\label{app:anomalies}

The purpose of this appendix is to check that Standard Model gauge dynamics does not explicitly break the Fricke symmetry.  We show that for the level-three assignment, mixed anomalies may be cancelled through $\tau$-dependent gauge-kinetic terms \cite{Feruglio:2023uof}, without introducing additional fermions.

We work in rigid $\mathcal N=1$ supersymmetry on spin manifolds, with gauge group $SU(3)_c\times SU(2)_L\times U(1)_y$ and integer hypercharge $y=6Y$. In particular, we assume that the gauge group is a direct product, rather than a quotient thereof \cite{Tong:2017oea}. All gauge-charged chiral fermions, including the higgsinos, contribute; the gauginos are modular neutral. Writing the character as $\epsilon=(-1)^p$, a canonically normalized chiral fermion transforms as
\bea
\psi\longmapsto\chi_\epsilon(\gamma)\left({j_\gamma(\tau)\over |j_\gamma(\tau)|}\right)^{-k}\psi~.
\eea
At the fixed points of $g=TW_3$ and $W_3$, its charges are $q_g=-k/2+3p\pmod6$ and $q_W=-k/2+p\pmod2$, respectively. For the gauge-charged fields they are
\bea\label{eq:anomalycharges}
\begin{array}{c|ccccccc}
 &Q&U^c&D^c&L&E^c&H_u&H_d\\\hline
q_g&(2,1,5)&(2,1,5)&(0,0,0)&(3,3,3)&(5,4,2)&0&3
\end{array}~,
\eea
with $q_W=q_g\bmod2$.

Define $\mathcal A_3(q)=\sum q\,2T_3$, $\mathcal A_2(q)=\sum q\,2T_2$, and $\mathcal A_{yy}(q)=\sum qy^2$, including multiplicities from the other gauge factors and using $2T(\mathbf{fund})=1$. Let $K_a$ and $P_a$ denote the same sums with $q$ replaced by $k$ and $p$, respectively. The results are
\bea\label{eq:mixedanomalychecks}
\begin{array}{c|rrrr}
a&\mathcal A_a(q_g)&\mathcal A_a(q_W)&K_a&P_a\\\hline
3&24&6&42&3\\
2&36&10&66&3\\
yy&1044&216&1260&54
\end{array}~.
\eea
The first two columns vanish modulo six and two, respectively. Thus gauge effects preserve the residual cyclic symmetries at the elliptic points. Away from these points, the automorphy factor produces a $\tau$-dependent measure phase controlled by the nonzero $K_a$.

To cancel this phase, we use the product of Dedekind eta functions $\Delta_3(\tau)=[\eta(\tau)\eta(3\tau)]^6$, which has weight six, is $T$ even and Fricke odd, and has no zeros in the upper half-plane.\footnote{Related eta products occur in string-loop gauge threshold corrections for heterotic orbifolds \cite{Bailin:1993fm}.} The functions
\bea\label{eq:anomalycounterterms}
\mathcal F_3=\Delta_3^7~,\hspace{0.4in} \mathcal F_2=\Delta_3^{11}~,\hspace{0.4 in} \mathcal F_y=\Delta_3^{210}~,
\eea
therefore have weights $K_a$ and characters $(-1)^{P_a}$. On a Fricke generator, the fermion-measure phase in each gauge sector is $-K_a\arg j_\gamma+\pi P_a$, multiplied by the integer gauge index. The variation of $\arg\mathcal F_a$ is $K_a\arg j_\gamma+\pi P_a$, so their sum is an integer multiple of $2\pi$. Holomorphic gauge-kinetic terms proportional to $\log\mathcal F_a$, with real coefficients fixed by the gauge-index normalization, then give rise to the desired anomaly cancellation, without changing the Yukawa textures or introducing a new source of CP violation. The branch of $\log\Delta_3$ that is real on the imaginary axis satisfies $\log\Delta_3(-\bar\tau)=\overline{\log\Delta_3(\tau)}$, making CP invariance explicit.

\section{The Fricke double-coset algebra}
\label{app:doublecosets}

In this final appendix, we derive the relation stated in (\ref{eq:doublecosetconnection}) and its action on the coefficient spaces. Throughout, we assume that $N=2,3$, weights are even, and modular groups are understood projectively.

To begin, define
\bea
\Gamma_N=\pm\Gamma(N)~,\hspace{0.4 in}
K_N=\Gamma_N\cap W_N\Gamma_NW_N^{-1}~.
\eea
An integral representative of an element of $K_N$ has $a=d=1\bmod N$, $b=0\bmod N$, and lower-left entry divisible by $N^2$, up to an overall sign.

At levels two and three, every element of $\Gamma_0(N)$ has a representative $\gamma=\left(\begin{smallmatrix}a&b\\Nc&d\end{smallmatrix}\right)$ with $a=d=1\bmod N$. Introduce the modular transformation $U_N:=W_NTW_N^{-1}=\left(\begin{smallmatrix}1&0\\-N&1\end{smallmatrix}\right)$. Writing $T$ and $U$ for the order-$N$ generators of the target group $\mathbb Z_N^T\times\mathbb Z_N^U$, define
\bea\label{eq:doublecosetquotientmap}
\pi(\gamma)=T^{[b]_N}U^{[-c]_N}\in\mathbb Z_N^T\times\mathbb Z_N^U~.
\eea
For $N=3$ the condition on $a,d$ selects the representative; for $N=2$ both signs give the same labels. Matrix multiplication adds the two exponents modulo $N$, so $\pi$ is a homomorphism with kernel $K_N$. The modular transformations $T$ and $U_N$ map to the two cyclic generators, proving surjectivity. Moreover,
\bea
W_N\begin{pmatrix}a&b\\Nc&d\end{pmatrix}W_N^{-1}
=\begin{pmatrix}d&-c\\-Nb&a\end{pmatrix}~.
\eea
Fricke conjugation exchanges the quotient labels. Hence $K_N$ is normal in the full Fricke group and
\bea\label{eq:doublecosetfinitequotient}
\Gamma_0^+(N)/K_N\simeq
G_N=(\mathbb Z_N^T\times\mathbb Z_N^U)\rtimes\mathbb Z_2^W~,
\hspace{0.4 in} \Gamma_N/K_N\simeq H=\langle U\rangle~.
\eea
The symbols $T,U,W$ on the right denote the finite images of the modular transformations $T,U_N,W_N$, respectively. In particular, $WTW^{-1}=U$, so $H$ is not normal. Taking double cosets then gives (\ref{eq:doublecosetconnection}), as claimed.

There are $N$ double cosets $HT^aH$ and one double coset $HWH$. To fix normalization, realize their algebra inside $\mathbb C[G_N]$ using
\bea
e_H={1\over N}\sum_{b=0}^{N-1}U^b~,\hspace{0.4 in}
 D_a=T^ae_H~,\hspace{0.4 in} X=Ne_HWe_H~.
\eea
The element $e_H$ is the identity of $e_H\mathbb C[G_N]e_H$. Since $We_HW^{-1}=N^{-1}\sum_aT^a$, multiplication gives
\bea\label{eq:doublecosetalgebra}
D_aD_b=D_{a+b}~,\hspace{0.4 in} D_aX=XD_a=X~,\hspace{0.4 in}
 X^2=N\sum_{a=0}^{N-1}D_a~.
\eea
This recovers the double-coset algebra described in Section \ref{sec:doublecosets}.

For its action on modular forms, the relevant decomposition is
\bea
\Gamma_NW_N\Gamma_N
 =\bigsqcup_{b=0}^{N-1}\Gamma_NT^bW_N~.
\eea
Thus on $M_k(\Gamma_N)$ the usual coset-sum operators are
\bea\label{eq:doublecosetslashaction}
D_af=f|_kT^a~,\hspace{0.4 in}
 Xf=\sum_{b=0}^{N-1}f|_k(T^bW_N)~,
\eea
which satisfy (\ref{eq:doublecosetalgebra}). On the subspace invariant under $T$, which is $M_k(\Gamma_0(N))$ for these levels, one has $Xf=N(f|_kW_N)$. Its two Fricke eigenspaces therefore give (\ref{eq:doublecosetcharacters}). 

To relate this action to the coefficient spaces in the main text, put $\omega=e^{2\pi i/N}$, $I_\pm^{(k)}=M_k^\pm(N)$, and $I_{R_r}^{(k)}=\{f\in M_k(\Gamma_N):f|_kT=\omega^r f\}$ for $r=1,\ldots,N-1$. The decomposition (\ref{eq:Frickecoefficientspaces}) then extends to
\bea\label{eq:residualsectordecomposition}
M_k(\Gamma_N)=I_+^{(k)}\oplus I_-^{(k)}\oplus\bigoplus_{r=1}^{N-1}I_{R_r}^{(k)}~.
\eea
On $I_{R_r}^{(k)}$, the scalar coset-sum action is $D_a=\omega^{ar}$ and $X=0$. These sectors contain the $H$-invariant components of the doublets described below.

Order-disorder doublets, such as the Ising $(\sigma,\mu)$ pair, involve both untwisted and defect-attached operators \cite{Bartsch:2023wvv}. For the coset construction above, analogous multiplets arise from two-dimensional representations $R_r$ of $G_N$. For $r=1,\ldots,N-1$, take $\Psi_r=(\sigma_r,\mu_r)^T$ with
\bea\label{eq:doublecosetmultiplets}
T\longmapsto\begin{pmatrix}\omega^r&0\\0&1\end{pmatrix}~,\hspace{0.3 in}
U\longmapsto\begin{pmatrix}1&0\\0&\omega^r\end{pmatrix}~,\hspace{0.3 in}
W\longmapsto\begin{pmatrix}0&1\\1&0\end{pmatrix}~.
\eea
For nonzero $f_r\in I_{R_r}^{(k)}$, the components of $(f_r,f_r|_kW_N)^T$ span a copy of $R_r$ in $M_k(K_N)$. Its second component has $U$ charge $\omega^r$ and thus lies outside the scalar space $M_k(\Gamma_N)$. Upon gauging $H=\langle U\rangle$, $\sigma_r$ remains local, while $\mu_r$ becomes an endpoint of an $H$ Wilson line. The duality defect descended from $W$ exchanges these components. For $N=3$, $r=1,2$ give conjugate doublets. We do not use these multiplets in the present paper.

\bibliographystyle{ytamsalpha}
\bibliography{bib}

\end{document}